\documentclass[pdflatex,sn-mathphys]{sn-jnl}% Math and Physical Sciences Reference Style
\usepackage{graphicx}% Include figure files
\usepackage{dcolumn}% Align table columns on decimal point
\usepackage{bm}% bold math
\usepackage{color} % For blue in-text comments and additions

\usepackage{amsmath,amssymb}
\jyear{2023}%

\theoremstyle{thmstyleone}%
\theoremstyle{thmstyletwo}%

\theoremstyle{thmstylethree}%

\begin{document}
%\preprint{NITEP 188}

\title[Intersections of ultracold atomic polarons and nuclear clusters]{Intersections of ultracold atomic polarons and nuclear clusters: How is a chart of nuclides modified in dilute neutron matter?}


\author*[1,2]{\fnm{Hiroyuki} \sur{Tajima}}\email{hiroyuki.tajima@phys.s.u-tokyo.ac.jp}

\author[3]{\fnm{Hajime} \sur{Moriya}}\email{moriya@nucl.sci.hokudai.ac.jp}

\author[4,5,2,3]{\fnm{Wataru} \sur{Horiuchi}}\email{whoriuchi@omu.ac.jp}

\author[6]{\fnm{Eiji} \sur{Nakano}}\email{e.nakano@kochi-u.ac.jp}

\author[6]{\fnm{Kei} \sur{Iida}}\email{iida@kochi-u.ac.jp}

\affil*[1]{\orgdiv{Department of Physics, Graduate School of Science}, \orgname{The University of Tokyo}, \orgaddress{%\street{7-3-1, Hongo, Bunkyo}, 
\city{Tokyo}, \postcode{113-0033}, \country{Japan}}}

\affil[2]{\orgname{RIKEN Nishina Center}, \orgaddress{%\street{2-1, Hirosawa},
\city{Wako}, \postcode{351-0198}, \country{Japan}}}

\affil[3]{\orgdiv{Graduate School of Science}, \orgname{Hokkaido University}, \orgaddress{%\street{N10W8}, 
\city{Sapporo}, \postcode{060-0810}, \country{Japan}}}

\affil[4]{\orgdiv{Department of Physics}, \orgname{Osaka Metropolitan University}, \orgaddress{%\street{3-3-138, Sugimoto, Sumiyoshi},
\city{Osaka}, \postcode{558-8585}, \country{Japan}}}

\affil[5]{\orgdiv{Nambu Yoichiro Institute of Theoretical and Experimental Physics (NITEP)}, \orgname{Osaka Metropolitan University}, \orgaddress{%\street{3-3-138, Sugimoto, Sumiyoshi}, 
\city{Osaka}, \postcode{558-8585}, \country{Japan}}}

\affil[6]{\orgdiv{Department of  Mathematics and Physics}, \orgname{Kochi University}, \orgaddress{%\street{2-5-1, Akebono}, 
\city{Kochi}, \postcode{780-8520}, \country{Japan}}}


\abstract{
Neutron star observations, as well as experiments on neutron-rich nuclei, used to motivate one to look at degenerate nuclear matter from its extreme, namely,
pure neutron matter.  As an important next step, impurities and clusters in dilute neutron matter have attracted special attention.
In this paper, we review 
in-medium properties of these objects on the basis of the physics of polarons, which have been
recently realized in ultracold atomic experiments. 
We discuss how such atomic and nuclear systems are related to each other in terms of polarons.
In addition to the interdisciplinary understanding of in-medium nuclear clusters, it is shown that
the quasiparticle energy of a single proton in neutron matter is associated 
with the symmetry energy, implying a novel route toward the nuclear equation of state 
from the neutron-rich side.
}




\maketitle

\section{Introduction}\label{sec1}
Low-temperature condensed matter physics burst into blossom in the early 20th century when metallic superconductivity and helium superfluidity were discovered.  Microscopically, the mechanism of superfluid helium II is still elusive, but the advent of the Bardeen-Cooper-Schrieffer (BCS) theory of superconductivity~\cite{PhysRev.108.1175} and Landau's theory of normal Fermi liquids~\cite{landau1959theory} in 1950's has led to the elucidation of the microscopic properties of various Fermi degenerate systems such as atomic nuclei and trapped cold Fermi atoms.
These systems are characterized by multiple degrees of freedom, while interactions between the constituent particles could give rise to order-disorder phase transitions, which might be accompanied by condensation of particle-particle and particle-hole pairs.

In the presence of such multiple degrees of freedom, we often encounter a situation in which the number of particles with a given state is far larger than that with another.  This kind of population imbalance could go to extremes where only a single minority particle, i.e., an impurity, is present in a system of majority particles, i.e., a medium.  When the impurity interacts with the medium but remains to be localized, the impurity behaves as a quasiparticle, of which the difference from a free particle in the energy dispersion contains information about the state of the medium
through dressing of the impurity by quantum fluctuations therein.
This kind of situation reminds us of
a polaron picture proposed by Landau and Pekar~\cite{landau1933electron,landau1948effective} for the microscopic description of a mobile electron in 
an ionic lattice in terms of a quasiparticle dressed by a cloud of phonons.
It is known that polaron effects on carriers play an important role in high-temperature superconductors~\cite{RevModPhys.78.17}.
Also, the polaron physics has been quantitatively examined in ultracold atomic gases, whose physical parameters are
experimentally tunable.
One of the most notable properties is the tunable interaction associated with a Feshbach resonance~\cite{RevModPhys.82.1225}.
Because of the mixing process between a scattering state of minority and majority atoms with different hyperfine states and a two-body bound state of the same pair of atoms, one can control the strength of the interatomic interaction by applying an external magnetic field to change the energy difference between those hyperfine states.
Such a technique enables us to investigate experimentally the interaction dependence of the polaron properties in a systematic manner~\cite{PhysRevLett.102.230402,PhysRevLett.103.170402,PhysRevA.84.011601,PhysRevLett.108.235302,koschorreck2012attractive,kohstall2012metastability,PhysRevLett.117.055301,PhysRevLett.117.055302,PhysRevLett.118.083602,PhysRevLett.122.093401,PhysRevLett.122.193604,PhysRevLett.125.133401,PhysRevA.103.053314,yan2020bose,duda2023transition} (for review, see, e.g.,  Refs.~\cite{chevy2010ultra,massignan2014polarons,schmidt2018universal,tajima2021polaron,scazza2022repulsive}).
Furthermore, as in the case of population-balanced superfluids in which an interaction-induced crossover from the weakly-coupled 
%Bardeen-Cooper-Schrieffer (BCS)-type superfluids
BCS state to the Bose-Einstein condensate (BEC) of tightly bound molecules has been studied extensively~\cite{chen2005bcs,zwerger2011bcs,doi:10.1146/annurev-conmatphys-031113-133829,STRINATI20181,OHASHI2020103739},
an interaction-induced transition or crossover from attractive polarons to dressed molecular clusters in population-imbalanced quantum gases has been of interest in the field of cold atoms~\cite{PhysRevX.10.041019}.
In this way, one can examine the polaron physics in various extreme conditions that cannot be realized in condensed matter systems.

It is thus natural to apply the polaron physics to nuclear systems (e.g., atomic nuclei and neutron stars) in which, in addition to multiple degrees of freedom associated with
spin, isospin, strangeness, etc., a liquid-gas phase transition in nuclear matter, i.e., a system of neutrons and protons, 
and an adsorption of neutrons at the liquid-gas interface are involved. 
Basically, atomic nuclei are quantum droplets of a liquid phase, which is induced by strong attraction responsible for the formation of deuterons and is characterized by a well-balanced neutron-proton mixture of a total number density close to the normal nuclear density, $\rho_0\approx0.16$ fm$^{-3}$; each of the two components is in a BCS superfluid state induced by an attractive central force in an isovector and spin-singlet channel at sufficiently low temperatures.  

Let us now consider neutron star matter, namely, the ground state of nuclear matter that is $\beta$ equilibrated and neutralized by electrons.  Well below $\rho_0$, the liquid phase corresponds to $^{56}$Fe nuclei, which are surrounded by a gas phase or, equivalently, the vacuum.  
With increasing density, the nuclei in equilibrium become more and more neutron-rich in such a way as to lower the electron chemical potential, and eventually the gas phase is filled with dripped neutrons out of the liquid phase.
Just below $\rho_0$, the system dissolves into uniform matter where the proton fraction can be as low as a few percent. At even higher densities,  the system may contain hyperons.  At these situations, which are encountered in neutron stars, protons and hyperons can be regarded as impurities in degenerate neutron matter~\cite{PhysRevC.47.1077,kutschera1994nuclear,PhysRevC.91.024327,PhysRevC.95.034309,RevModPhys.88.035004}.
At finite temperatures, the system below $\rho_0$ is in nuclear statistical equilibrium, i.e., composed of various kinds of nuclei according to the Boltzmann factor.  Then, the system contains impurity-like light clusters such as alpha particles, $^3$He, and tritons in extremely
neutron-rich environments, which are relevant to binary neutron-star mergers~\cite{lalit2019dense}.  
Heavy nuclei present are sufficiently neutron-rich to have a neutron skin region in which neutrons adsorb onto the nuclear surface, i.e., the liquid-gas interface.  This specific region may also contain light clusters even at zero temperature; in recent experiments of heavy neutron-rich nuclei such as stable tin isotopes, the emergence of impurity-like alpha clusters near the surface has been revealed~\cite{tanaka2021formation}.
%The concept of polarons is not limited to condensed matter or cold atomic systems.
%One can apply the polaron description to impurity problems in nuclear physics.
%Indeed, impurity-like light clusters in neutron-rich environments are relevant to 
We remark that the system in stellar collapse is similar to that in binary neutron star mergers, but is more proton-rich in the liquid phase because of neutrinos trapped in supernova cores and hence is more dilute and less degenerate in the gas phase~\cite{JANKA200738}. 
These \textit{nuclear impurity problems} are challenging because impurity dressing and clustering by the neutron medium
cannot be addressed by conventional mean-field approaches.

The advantage of the polaronic approach to such impurity problems is its interdisciplinary 
nature that can be described by such quantities as the Fermi momentum of the medium and the scattering length between the impurity and 
the medium particle, but is almost independent of the microscopic details of the medium.
Since theoretically predicted polaron properties
can be tested via the comparison with cold atomic experiments, from such well-tested predictions and, if necessary, their modifications, e.g., from a nonzero effective range of the impurity-medium interaction, one can examine polaronic effects in astrophysical conditions that are not accessible from nuclear experiments. 

In this article, we review such intersections between ultracold atomic and nuclear impurity problems.
The first question we would like to address here is
how 
the many-body physics associated with polaron formation is visible in ultracold atomic experiments with various settings where statistics of medium atoms and interspecies interactions can be controlled.
Second, we present theoretical studies of nuclear impurity problems that are based on the polaron picture reinforced by the recent ultracold atomic experiments.
In particular,
we focus on an alpha particle and a proton immersed in dilute neutron matter, which are analogous to ultracold atomic polarons and 
are relevant both in astrophysical environments and in nuclear experiments.

This article is organized as follows.
In Sec.~\ref{sec:2},
we first overview the recent development of ultracold atomic polarons.
Two classes of polarons, that is, Fermi and Bose polarons are reviewed.
In addition, we mention a $P$-wave Feshbach resonance, which is associated with neutron-alpha scattering discussed in the next section.
In Sec.~\ref{sec:3}, based on the notion of ultracold atomic polarons, we argue how the chart of nuclides, which normally depicts all known bound nuclides in vacuum, can be modified in dilute neutron matter.
In particular, we pick up polaronic alpha particles and protons as important examples.
After introducing the basic setting of polaronic alpha particles,
we discuss the possible formation of in-medium two- and three-alpha bound states due to polaronic effects.
Also, we show how the polaronic proton energy is related to the nuclear symmetry energy and then how clustering
of polaronic protons 
occurs in neutron-rich matter.
In Sec.~\ref{sec:4}, we give a conclusion.

%The Introduction section, of referenced text \cite{bib1} expands on the background of the work (some overlap with the Abstract is acceptable). The introduction should not include subheadings.

%Springer Nature does not impose a strict layout as standard however authors are advised to check the individual requirements for the journal they are planning to submit to as there may be journal-level preferences. When preparing your text please also be aware that some stylistic choices are not supported in full text XML (publication version), including coloured font. These will not be replicated in the typeset article if it is accepted. 

\section{Ultracold atomic polarons}
\label{sec:2}
%\begin{widetext}
\begin{figure}[t]
    \centering
    \includegraphics[width=12cm]{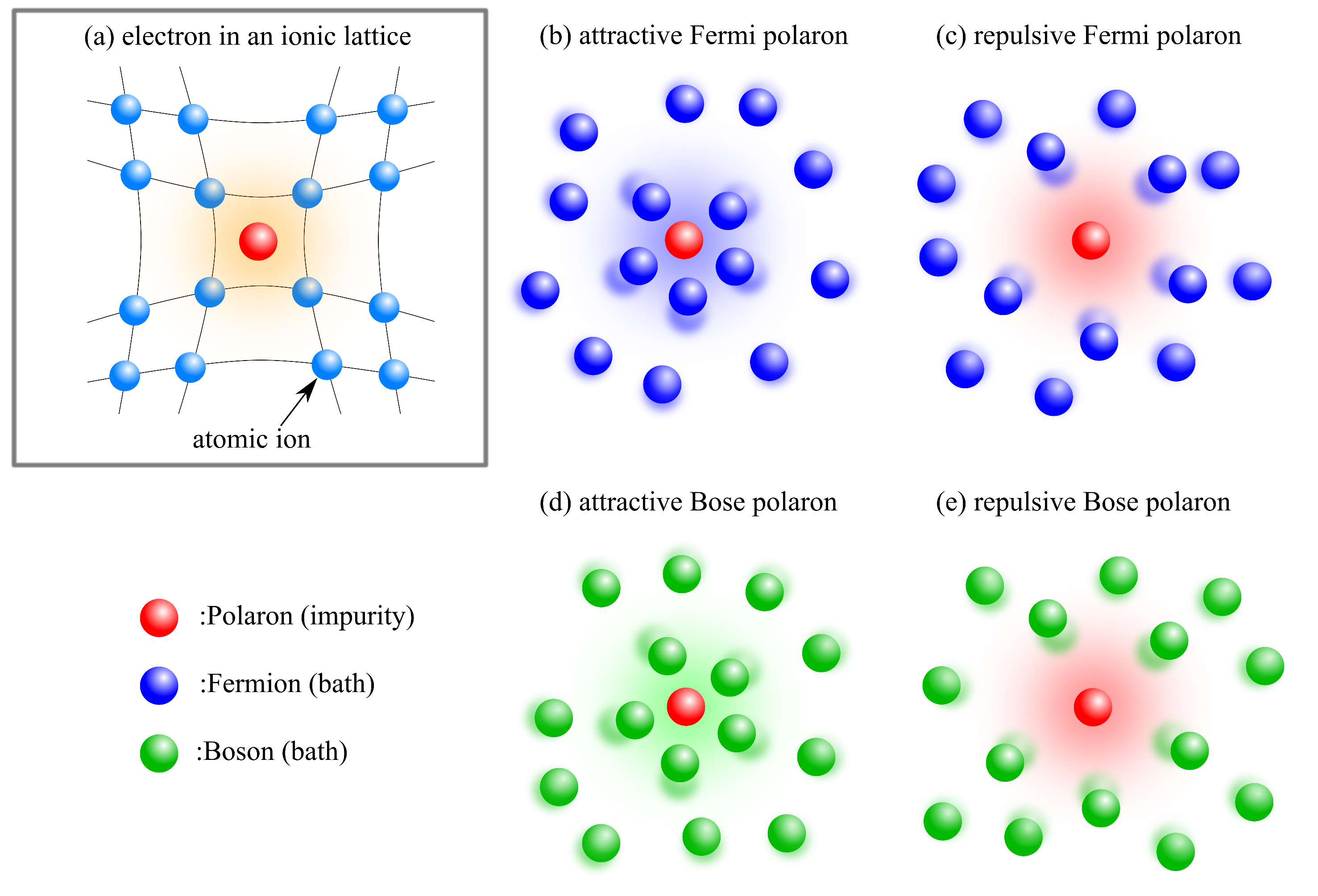}
    \caption{Schematics of polarons in condensed matter and in cold atom systems.
    (a) An electron moving in the lattice of atomic ions;
    (b) an attractive Fermi polaron;
    (c) a repulsive Fermi polaron;
    (d) an attractive Bose polaron;
    (e) a repulsive Bose polaron (see the text for more details).
    }
    \label{fig:1}
\end{figure}
%\end{widetext}
In this section, we review recent progress in the polaron problems involving ultracold atomic gases.
We take the system volume $\mathcal{V}$ to be unity and $\hbar=k_{\rm B}=1$ for convenience.
While the polaron picture was originally applied to an electron moving in an ionic lattice 
as shown in Fig.~\ref{fig:1}, ultracold atomic polarons are realized by preparing a two-component mixture with large population imbalance.
Generally, such a system can be described by the Hamiltonian,
\begin{align}
\label{eq:Hamiltonian}
    H&=\sum_{\bm{k}}\varepsilon_{\bm{k},{\rm i}}
    c_{\bm{k},{\rm i}}^\dag c_{\bm{k},{\rm i}}
    +\sum_{\bm{k}}\varepsilon_{\bm{k},{\rm m}}
    c_{\bm{k},{\rm m}}^\dag c_{\bm{k},{\rm m}}\cr
    &+\frac{1}{2}
    \sum_{\bm{k},\bm{k}',\bm{P}}
    U_{\rm ii}(\bm{k},\bm{k}')
    c_{\bm{k}+\bm{P}/2,{\rm i}}^\dag
    c_{-\bm{k}+\bm{P}/2,{\rm i}}^\dag
    c_{-\bm{k}'+\bm{P}/2,{\rm i}}
    c_{\bm{k}'+\bm{P}/2,{\rm i}}\cr
    &+\frac{1}{2}
    \sum_{\bm{k},\bm{k}',\bm{P}}
    U_{\rm mm}(\bm{k},\bm{k}')
    c_{\bm{k}+\bm{P}/2,{\rm m}}^\dag
    c_{-\bm{k}+\bm{P}/2,{\rm m}}^\dag
    c_{-\bm{k}'+\bm{P}/2,{\rm m}}
    c_{\bm{k}'+\bm{P}/2,{\rm m}}\cr
    &+
    \sum_{\bm{k},\bm{k}',\bm{P}}
    U_{\rm im}(\bm{k},\bm{k}')
    c_{\bm{k}+\bm{P}/2,{\rm i}}^\dag
    c_{-\bm{k}+\bm{P}/2,{\rm m}}^\dag
    c_{-\bm{k}'+\bm{P}/2,{\rm m}}
    c_{\bm{k}'+\bm{P}/2,{\rm i}},
\end{align}
where $\varepsilon_{\bm{k},{\rm i}}$ and $\varepsilon_{\bm{k},{\rm m}}$
are the kinetic energies of atoms with different hyperfine states denoted by the labels``i" and ``m," respectively,
and with momentum $\bm{k}$, which
is a quantum number since we focus on a homogeneous system.
$a_{\bm{k}}^{(\dag)}$ and $b_{\bm{k}}^{(\dag)}$ are the annihilation (creation) operators of atomic species ``i" and ``m," respectively. 
The third and fourth terms in Eq.~\eqref{eq:Hamiltonian} represent the intra-species interaction where the coupling strengths $U_{\rm ii}(\bm{k},\bm{k}')$ and $U_{\rm mm}(\bm{k},\bm{k}')$ do not depend on the two-particle center-of-mass momentum $\bm{P}$
because of the translational invariance.
The inter-species interaction given by the last term in Eq.~\eqref{eq:Hamiltonian} plays a crucial role in determining the polaron 
properties.  Again, its coupling strength $U_{\rm im}(\bm{k},\bm{k}')$ does not depend on $\bm{P}$.

The number density of each component reads
\begin{align}
    \rho_{{\rm i}}=\sum_{\bm{k}}\langle c_{\bm{k},{\rm i}}^\dag c_{\bm{k},{\rm i}}\rangle,
\end{align}
\begin{align}
    \rho_{{\rm m}}=\sum_{\bm{k}}\langle c_{\bm{k},{\rm m}}^\dag c_{\bm{k},{\rm m}}\rangle,
\end{align}
where $\langle\cdots\rangle$ denotes the equilibrium average.
Hereafter, the ``i"- and ``m"-components are regarded as impurity (minority) and medium (majority) atoms, respectively.  
In other words, we consider the case in which $\rho_{\rm i} \ll \rho_{\rm m}$.
While a single impurity is usually considered in theory,
in actual experiments, the minority component may well have a macroscopic number $N_{\rm i}\gg 1$.
We note that the atomic number is associated with the density as 
$N_{\rm i,m}=\rho_{\rm i,m}\mathcal{V}$.

In cold atomic systems, polaronic excitations can be probed via radio-frequency (RF) spectroscopy~\cite{torma2014quantum}.
There are mainly two kinds of RF responses called the injection-type~\cite{scazza2022repulsive} 
\begin{align}
    I_{\rm in}(\omega)=2\pi\Omega_{\rm R}^2\sum_{\bm{k}}
    f(\varepsilon_{\bm{k},{\rm ref.}})A_{\rm i}(\bm{k},\varepsilon_{\bm{k},{\rm i}}+\omega),
\end{align}
and the ejection-type
\begin{align}
    I_{\rm ej}(\omega)=2\pi\Omega_{\rm R}^2
    \sum_{\bm{k}}f(\varepsilon_{\bm{k},{\rm i}}-\omega)
    A_{\rm i}(\bm{k},\varepsilon_{\bm{k},{\rm i}}-\omega),
\end{align}
where $\varepsilon_{\bm{k},{\rm ref.}}$ is the kinetic energy of the reference state with a different hyperfine state,
$f(x)$ is the distribution function,
$\Omega_{\rm R}$ is the Rabi coupling, and $\omega$ is the RF frequency.
Here we ignore initial and final state interactions for the injection and ejection RF spectroscopies, respectively.
In this way, the impurity spectral function $A_{\rm i}(\bm{k},\omega)$ can be probed via the RF spectroscopy.
$A_{\rm i}(\bm{k},\omega)$ gives information on the polaronic excitation.
For uniform systems with $\varepsilon_{\bm{k},{\rm i}}=k^2/2M$ where $M$ is the mass of an impurity atom,
the retarded propagator of a polaron with a positive infinitesimal $\delta$ is given by
\begin{align}
    G_{\rm i}(\bm{k},\omega)&=\frac{1}{\omega+i\delta-\varepsilon_{\bm{k},{\rm i}}-\Sigma_{\rm i}(\bm{k},\omega)}\cr
    &\simeq
    \frac{Z}{\omega+i\delta-\frac{k^2}{2M^*}-E_{\rm P}+i\Gamma/2},
\end{align}
where
the polaron energy $E_{\rm P}$, the polaron residue $Z$, the decay rate $\Gamma$, and the effective mass $M^*$ can be extracted from the impurity self-energy $\Sigma_{\rm i}(\bm{k},\omega)$ as
\begin{align}
    E_{\rm P}={\rm Re}\Sigma_{\rm i}(\bm{0},E_{\rm P}),
\end{align}
\begin{align}
    Z=\left[1-{\rm Re}\left(\frac{\partial \Sigma_{\rm i}(\bm{0},\omega)}{\partial \omega}\right)_{\omega=E_{\rm P}}\right]^{-1},
\end{align}
\begin{align}
    \frac{M}{M^*}=Z\left[1+M{\rm Re}\left(\frac{\partial^2 \Sigma_{\rm i}(\bm{k},E_{P})}{\partial k^2}\right)_{\bm{k}=\bm{0}}\right],
\end{align}
\begin{align}
    \Gamma=-2Z{\rm Im}\Sigma_{\rm i}(\bm{0},E_{P}).
\end{align}
These quantities can be extracted from the RF spectra through the relation $A_{\rm i}(\bm{k},\omega)=-\frac{1}{\pi}{\rm Im}G_{\rm i}(\bm{k},\omega)$.

At zero temperature ($T=0$),
$E_{\rm P}$ is directly related to the thermodynamic quantities such as the impurity chemical potential $\mu_{\rm i}$ and the internal energy density $E$.
Since
$E_{\rm P}$ and $\mu_{\rm i}=\frac{\partial E}{\partial \rho_{\rm i}}$ are defined as the energy change induced by adding an impurity to the system,
 $\mu_{\rm i}$ is equivalent to $E_{\rm P}$ at $T=0$.
In the limit of $\rho_{\rm i}/\rho_{\rm m}\rightarrow 0$, therefore, one can write
\begin{align}
\label{eq:eg}
    E\simeq E_{\rm m}+E_{\rm P}\rho_{\rm i},
\end{align}
where $E_{\rm m}$ is the ground-state energy density of a pure system 
without impurities.
In this way, one can see a clear relation between the thermodynamic quantities and the polaron properties at certain conditions.

Ultracold atomic polarons are known to have different single-particle properties in a manner that is dependent on the quantum statistics 
of the majority component. Accordingly, there are two classes as depicted in Fig.~\ref{fig:1}.
A Fermi (Bose) polaron corresponds to an impurity atom immersed in a degenerate Fermi (Bose) atomic gas; 
for further clarity, "attractive" or "repulsive" is often used to classify the sign of the impurity-medium interaction $U_{\rm im}$.

\subsection{Fermi polarons}
\label{subsec:FP}
We start with a Fermi polaron, i.e., 
the operator $c_{\bm{k},{\rm m}}^{(\dag)}$ of the majority component obeys 
the anti-commutation relations $\{c_{\bm{k},{\rm m}},c_{\bm{k}',{\rm m}}^\dag\}=\delta_{\bm{k},\bm{k}'}$,
$\{c_{\bm{k},{\rm m}},c_{\bm{k}',{\rm m}}\}=\{c_{\bm{k},{\rm m}}^\dag,c_{\bm{k}',{\rm m}}^\dag\}=0$.
%On the other hand, 
Note that the impurity operator $c_{\bm{k},{\rm i}}^{(\dag)}$ can be either fermionic or bosonic. 
This difference is not important in the single-impurity limit ($\rho_{\rm i}/\rho_{\rm m}\rightarrow 0$), but as
the number of impurities increases, it becomes significant as the medium-induced polaron-polaron interaction generally depends on the quantum statistics of impurities even for a pair of different internal states.

Experimentally, as explained in the introduction, the impurity-medium interaction is controlled by Feshbach resonances~\cite{RevModPhys.82.1225}.
In particular, the polaron systems with the tunable $S$-wave interaction have been studied extensively.
In such a case, the impurity-medium interaction can be characterized by the $S$-wave scattering length $a_{\rm im}$,
which helps to connect between the ultracold atomic and nuclear impurity problems as will be seen later.
While the case of the attractive impurity-bath interaction is called ``attractive Fermi polarons,"
the repulsive case is called ``repulsive Fermi polarons"~\cite{chevy2010ultra,massignan2014polarons,schmidt2018universal,tajima2021polaron,scazza2022repulsive}.
For atomic Fermi gases with population imbalance,
$U_{\rm mm}(\bm{k},\bm{k}')$ is usually negligible because the majority component consists of fermions in the same hyperfine state
and hence has the low-energy $S$-wave scattering 
suppressed by the Pauli-blocking effect.
The effect of $U_{\rm ii}(\bm{k},\bm{k}')$ is also negligible basically when the impurity density is sufficiently small.
Anyway, Feshbach resonances are used for enhancing $U_{\rm im}(\bm{k},\bm{k}')$ alone, while $U_{\rm mm}(\bm{k},\bm{k}')$ and 
$U_{\rm ii}(\bm{k},\bm{k}')$ remain off-resonant residual interactions (except for the presence of the overlapped Feshbach resonances).
In nuclear impurity problems, however, we remark that
$U_{\rm ii}(\bm{k},\bm{k}')$ can play a crucial role as will be shown in Sec.~\ref{sec:3}.

At negative $a_{\rm im}$, an attractive Fermi polaron, whose quasiparticle properties such as the polaron energy, the decay width, the residue, and the effective mass have been measured experimentally~\cite{PhysRevLett.102.230402,PhysRevLett.118.083602},
is quite stable.  Such stability is maintained even around the unitary limit ($a_{\rm im}\rightarrow-\infty$).
If the attractive interaction is sufficiently strong to go beyond the unitary limit, i.e., $a_{\rm im}^{-1}$ goes above zero, a dressed molecular state of an impurity atom and a medium atom can be more stable than the attractive polaron. 
This structural change, which is called a polaron-to-molecule transition or crossover~\cite{PhysRevA.80.053605}, is expected to occur at $k_{\rm F}a_{\rm im}\simeq 1$, where $k_{\rm F}$ is the Fermi momentum of the medium.
While it was originally predicted as a first-order phase transition at $T=0$~\cite{PhysRevA.80.053605} and numerically confirmed by the diagrammatic Monte Carlo  method~\cite{PhysRevB.77.020408,PhysRevB.77.125101},
recent experiments indicate that such a transition becomes a smooth crossover under the influence of
non-zero temperature and impurity densities~\cite{PhysRevX.10.041019}.

Theoretical studies on the Fermi polaron problems have gone across different spatial dimensions, while having made full use of various calculation methodologies. 
Typically, such studies are based on the variational approach developed by Chevy~\cite{PhysRevA.74.063628}, the diagrammatic approach involving the in-medium $T$-matrix~\cite{PhysRevLett.98.180402,PhysRevA.77.031601,PhysRevA.78.031602,PhysRevLett.105.020403,PhysRevA.84.033607,PhysRevA.98.013626,Tajima_2018,PhysRevA.99.063606,PhysRevB.107.054505},
the renormalization group~\cite{PhysRevLett.100.140407,PhysRevA.83.063620,PhysRevA.95.013612,PhysRevA.105.043303},
the diffusion Monte Carlo method~\cite{PhysRevLett.97.200403,PhysRevLett.100.030401,PhysRevA.104.043313},
the diagrammatic Monte Carlo method~\cite{PhysRevB.77.020408,PhysRevB.89.085119,PhysRevB.91.144507,PhysRevA.94.051605,PhysRevB.101.045134},
and the lattice action~\cite{PhysRevLett.115.185301,hildenbrand2022lattice,PhysRevResearch.3.033180}.
The obtained results for the quasiparticle properties of Fermi polarons are known to fairly well reproduce the available experimental results~\cite{scazza2022repulsive}.

In the case of attractive Fermi polarons in which impurities are fermions embedded in a three-dimensional medium, the ground-state energy density of the whole system can be expanded in series of $\rho_{\rm i}/\rho_{\rm m}$ with the coefficients determined by the polaron properties as~\cite{PhysRevLett.97.200403}
\begin{align}
\label{eq:Epol}
    E=
    E_{\rm FG}
    \left[1+\frac{5}{3}\frac{E_{\rm P}}{E_{\rm F}}\left(\frac{\rho_{\rm i}}{\rho_{\rm m}}\right)+\frac{M}{M^*}\left(\frac{\rho_{\rm i}}{\rho_{\rm m}}\right)^{\frac{5}{3}}+\cdots\right],
\end{align}
where $E_{\rm FG}=\frac{3}{5}\rho_{\rm m}E_{\rm F}$ and $E_{\rm F}$ are the ground-state energy density and the Fermi energy, respectively, of the majority Fermi gas.
The first two terms in Eq.~\eqref{eq:Epol} 
just correspond to Eq.~\eqref{eq:eg}, the form of which holds for whatever statistics majority and minority atoms obey, while
the term proportional to $M/M^*$ originates from the Fermi degeneracy of impurity atoms and thus would be absent for bosonic impurities. 
The energy and effective mass of an attractive Fermi polaron 
can be deduced from observations of the lowest compression mode of 
the system in the unitary limit~\cite{PhysRevLett.103.170402,Sommer_2011};
the corresponding eigen-frequency is given by
\begin{align}
    \omega^*=\omega\sqrt{\left(1-\frac{E_{\rm P}}{E_{\rm F}}\right)\frac{M}{M^*}},
\end{align}
where $\omega$ is the trap frequency along the direction of the resultant dipole oscillation of the two components.
Such a collective mode is similar to giant dipole resonances in neutron-rich nuclei~\cite{BERTULANI1993281} where protons can be regarded as impurities.
%in neutron-rich matter.

\subsection{Bose polarons}
Let us now turn to a Bose polaron, i.e., an impurity atom immersed in an atomic BEC (see also Fig.~\ref{fig:1}). 
As in the case of Fermi polarons,
both attractive and repulsive Bose polarons have been studied extensively~\cite{PhysRevLett.117.055301,PhysRevLett.117.055302,yan2020bose,duda2023transition}.
Interestingly, various non-trivial phenomena  
such as collective mode excitations and few-body clustering are involved in Bose polarons.
Several theoretical methods including variational and diagrammatic approaches
have been exploited to understand such Bose polaron physics~\cite{Devreese_2009,PhysRevA.90.013618,vlietinck2015diagrammatic,PhysRevA.88.053632,PhysRevB.93.205144,PhysRevLett.115.125302,PhysRevA.96.013607,PhysRevLett.122.183001,PhysRevA.99.063607,PhysRevA.100.023624,Mistakidis_2020}.
In particular, possible formation of two-impurity bound states and bipolarons with the help of phonon-mediated interactions has been studied theoretically~\cite{PhysRevA.88.013613,naidon2018two,PhysRevLett.121.013401,PhysRevLett.121.080405,PhysRevResearch.2.023154,PhysRevLett.127.103401}.

In this article, we do not go into details about applications of the Bose-polaron picture to nuclear physics.
A possible candidate would be a nucleon immersed in a cloud of mesons~\cite{PhysRevD.34.2112,PhysRevD.53.3337,PhysRevD.53.3354}.
Another would be an impurity in dilute alpha matter where each alpha particle, if stable, could be regarded as a building block.
%a fundamental degree of freedom.
Theoretical study on such Bose polarons in alpha matter will be reported elsewhere~\cite{Tajima2023alpha}.

\subsection{$P$-wave resonance}
In Sec.\ \ref{subsec:FP}, we confine ourselves to cold atomic Fermi polarons formed by the $S$-wave impurity-medium interaction.  This is reasonable given experimental realizations in which a trapped gas of the majority atoms is sufficiently dilute for the interparticle spacing to be far larger than the effective range of the interaction.  In principle, however,
$P$-wave Feshbach resonances can be used for controlling the impurity-medium interaction.
So far, there have been only a few theoretical studies on the resultant $P$-wave Fermi polarons~\cite{PhysRevLett.109.075302,PhysRevA.100.062712}.

Here we present a basic theoretical model for the $P$-wave resonance, which is relevant to not only cold atomic but also nuclear polaron problems.
For the $P$-wave Feshbach resonance, instead of the single-channel model used in Eq.~\eqref{eq:Hamiltonian},
it is convenient to start from the so-called two-channel model in which
the impurity-medium interaction can be written as~\cite{gurarie2007resonantly}
\begin{align}
\label{eq:Pwave}
    H_{\rm FR}=\sum_{\bm{k},\bm{P},L,L_z}
    \left(
    g_{\bm{k}}C_{\bm{P},L,L_z}^\dag
    c_{\bm{k}+\bm{P}/2,{\rm m}}
    c_{-\bm{k}+\bm{P}/2,{\rm i}}
    +    g_{\bm{k}}^*
    c_{\bm{k}+\bm{P}/2,{\rm m}}^\dag
    c_{-\bm{k}+\bm{P}/2,{\rm i}}^\dag
    C_{\bm{P},L,L_z}
    \right),
\end{align}
where $C_{\bm{P},L,L_z}^{(\dag)}$ is the annihilation (creation) operator of a closed-channel Feshbach molecule with momentum $\bm{P}$, orbital angular-momentum quantum number $L$, and its $z$-component $L_z$.
For simplicity, we ignore off-resonant background interactions. 
Moreover, we need to consider the kinetic term of closed-channel Feshbach molecules given by~\cite{gurarie2007resonantly}
\begin{align}
\label{eq:FM}
    H_{\rm FM}=\sum_{\bm{P},L,L_z}
    \left(\varepsilon_{\bm{P},{\rm FM}}+\nu_{L,L_z}\right)
    C_{\bm{P},L,L_z}^\dag C_{\bm{P},L,L_z},
\end{align}
where $\varepsilon_{\bm{P},{\rm FM}}$ is the kinetic energy of a closed-channel Feshbach molecule of momentum $\bm{P}$. 
For the $P$-wave channel, i.e., $L=1$, the Feshbach coupling can be anisotropic as
$g_{\bm{k}}\propto Y_{L_z}^{1}(\hat{\bm{k}})$,    
where $Y_{L_z}^{1}(\hat{\bm{k}})$ is the spherical harmonics with $L_z=0,\pm 1$.
The form of $g_{\bm{k}}$ can be determined in such a way as to reproduce the low $\bm{k}$ behavior of the $P$-wave phase shift $\delta_P(\bm{k})$ as given by
\begin{align}
    k^3\cot\delta_P(\bm{k})=-\frac{1}{v_P}+\frac{1}{2}r_P k^2+O(k^3),
\end{align}
where $v_P$ and $r_P$ are the $P$-wave scattering volume and ``effective range," respectively.

%There are few previous studies about atomic polarons with $P$-wave impurity-bath interaction, which are called $P$-wave polarons~\cite{PhysRevLett.109.075302,PhysRevA.100.062712}. In particular, 
In Ref.~\cite{PhysRevLett.109.075302},
it was reported that $P$-wave Fermi polarons exhibit distinct features from the $S$-wave counterpart;
in a magnetic field, there appears a third polaron branch in the excitation spectrum, 
in addition to the usual attractive and repulsive ones, which in turn exhibit an anisotropic dispersion of the impurity characterized by different effective masses perpendicular and parallel to the magnetic field.  However, there remain intriguing similarities between the $P$-wave and $S$-wave cases.  The characteristics of the $P$-wave polaron-to-molecule transition are theoretically found to be similar to the case near a narrow $S$-wave Feshbach resonance~\cite{kohstall2012metastability,Massignan_2012}.  In Ref.~\cite{PhysRevA.100.062712}, the same kind of transition was considered for an impurity interacting with a one-dimensional majority Fermi gas via a narrow $P$-wave Feshbach resonance.
Interestingly, $P$-wave resonant scattering in one dimension has a close similarity with the $S$-wave one in three dimensions~\cite{PhysRevA.94.043636,PhysRevA.104.023319,PhysRevB.105.064508,PhysRevA.107.033331}.
While $P$-wave polarons have not been realized in experiments,
known $P$-wave Feshbach resonances could be utilized to control the impurity-medium $P$-wave interaction in future experiments.

\section{Applications to nuclear impurity problems}
\label{sec:3}
\begin{figure}[t]
    \centering
    \includegraphics[width=12cm]{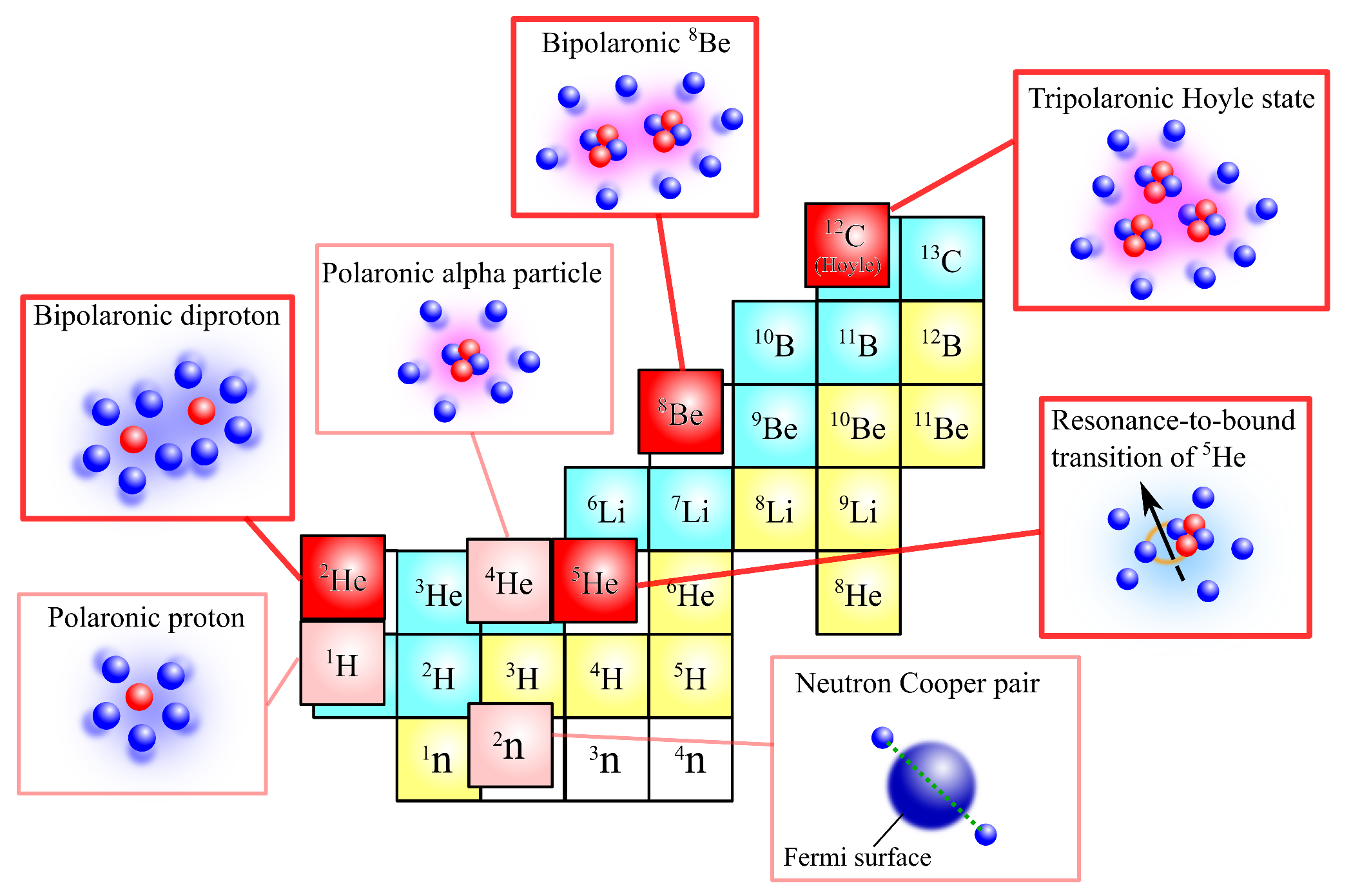}
    \caption{Modified nuclear chart in the presence of a neutron gas. The light red colored nuclei (i.e., $^1$H, $^{4}$He, and $^2$n) are 
    %well-established objects including stable
    quasiparticles (polaronic proton and alpha particle) immersed in neutron matter and a dineutron, namely, a neutron Cooper pair. The thick red colored nuclei (i.e., $^{2}$He, $^5$He, $^{8}$Be, and the $^{12}$C Hoyle state) are 
    are known to be unbound in vacuum, but are predicted to be bound in neutron medium~\cite{PhysRevC.104.065801,PhysRevC.106.045807,tajima2023polaronic}.}
    \label{fig:2}
\end{figure}

We are now in a position to apply the polaron physics developed for cold atoms experimentally and theoretically to nuclear impurity problems.
Such applications might be concisely summarized by drawing a chart of nuclides embedded in a gas of neutrons,
as shown in Fig.~\ref{fig:2}.
In vacuum, the $^2$He, $^5$He, $^8$Be ground states, and the Hoyle state (the 2nd excited state of $^{12}$C) are known to be unbound.
It is, however, suggested that in a cold neutron gas, as encountered in 
neutron stars 
as well as neutron-rich nuclei, $^2$He, $^5$He, $^8$Be, and the Hoyle state could turn into 
a bound state of two polaronic protons, i.e., a diproton, a Feshbach molecule of an alpha particle and a neutron,  
a bound state of two polaronic alpha particles, and a bound state of three polaronic alpha particles, respectively.
In the following, we will briefly discuss how such binding could occur.
%\end{widetext}

\subsection{Polaronic alpha particles}
% interaction has to be energy independent?
In the low-energy limit of a relative motion of a neutron-alpha two-particle system, 
%the neutron-alpha interaction exhibits 
the $S$-wave scattering dominates the cross section.  
The empirical $S$-wave neutron-alpha scattering length $a_{\rm n\alpha}$ and effective range $r_{\rm n\alpha}$ 
are known as $a_{\rm n\alpha}=2.64$~fm and $r_{\rm n\alpha}=1.43$~fm~\cite{kanada1979microscopic}. 
In this channel, repulsion is induced by
the Pauli-blocking effect between the free neutron and two valence neutrons in the alpha particle.
Note that the
alpha particle can be regarded as a point-like particle 
throughout the scattering because of its relatively strong binding.

For an impurity alpha particle that is at rest or slowly moving in sufficiently dilute neutron matter to ensure $k_F r_{\rm n\alpha}\ll1$, therefore,
one may regard it as an $S$-wave repulsive Fermi  polaron discussed in Sec.\ \ref{subsec:FP}.
The quasiparticle properties of such an alpha particle have been
investigated by allowing for a single neutron particle-hole pair excitation a la Chevy's variational ansatz~\cite{PhysRevC.102.055802}.
The repulsive neutron-alpha interaction leads to a positive (but small) polaron energy, i.e., $0<E_{\rm P}<E_{\rm F}$.
This indicates that in such a polaron, the alpha particle repels surrounding neutrons, while the repulsive interaction is so weak that the alpha particle and the neutron medium are still miscible.
This alpha particle has its effective mass enhanced by carrying a dressing cloud of neutron holes with it.
Indeed, similar effective mass corrections have been observed for repulsive Fermi polarons in $^6$Li atomic experiments~\cite{PhysRevLett.118.083602}.

Remarkably,
this enhancement of the effective mass, together with the Ruderman-Kittel-Kasuya-Yosida (RKKY) type medium-induced interaction between impurities~\cite{PhysRev.96.99,10.1143/PTP.16.45,PhysRev.106.893}, is crucial
for the formation of multi-polaron bound states.
In Ref.~\cite{PhysRevC.104.065801},
the present authors showed that unbound alpha clusters in vacuum can be bound states in medium due to such polaronic many-body effects.
In particular, as schematically summarized in the in-medium nuclear chart (i.e., Fig.~\ref{fig:2}), $^8$Be, which is, in vacuum, an unbound nucleus 
that radiatively decays into two alpha particles, 
is found to turn into a bound state because of the neutron-mediated short-range attraction between two alpha particles as well as the increased effective mass of each alpha particle.
Furthermore, the Hoyle state~\cite{hoyle1954nuclear}, which is  essentially a resonant
three alpha-particle state of $^{12}$C and plays a crucial role in nucleosynthesis, is also found to become a bound state through the competition among neutron-mediated two-body attraction and three-body repulsion as well as the effective mass correction.
These in-medium bound states are analogous to atomic multi-polarons~\cite{Devreese_2009}.
While clear empirical evidence for the presence of bi- and tri-polarons has yet to be obtained
in both atomic and nuclear fields,
it would be worthwhile keep the in-medium nuclear chart updated.

At higher neutron densities,
the above picture of a neutron-alpha mixture that interacts with each other via the 
$S$-wave interaction becomes gradually insufficient for several reasons.
The most important reason is that
the $^5$He ground-state narrow resonance, which is, in vacuum, known to occur in the $P$-wave channel of total angular momentum $J=3/2$,
would be relevant in the relatively low-density regime.
Given that a $P$-wave neutron-alpha interaction of some form is responsible for
the $^5$He resonance~\cite{PhysRevC.106.045807},
one can theoretically relate the $J^{\pi}=3/2^{-}$ $P$-wave scattering amplitude and the $^5$He resonance energy through the analogue model of the $P$-wave Feshbach resonance shown in Eqs.~\eqref{eq:Pwave} and \eqref{eq:FM}.
Here we ignore the insignificant $J^{\pi}=1/2^{-}$ $P$-wave channel for simplicity. 
In a sufficiently dilute neutron gas,
the $^5$He resonant state corresponds to the closed-channel Feshbach molecule described by Eq.~\eqref{eq:FM} in the BCS-like weak coupling regime where the $P$-wave scattering volume is negative and hence the closed-channel molecular energy level is higher than the bottom of the two-body continuum.
If the center-of-mass neutron kinetic energy corresponding to
the Fermi energy $E_{\rm F}$ of the medium (a neutron gas in the present case) reaches the resonance energy of about 0.9 MeV,
this resonance can turn into a bound state because the decay process of the molecule to an alpha particle and a neutron
is prohibited by the Pauli-blocking effect.
We remark that
this resonance-to-bound transition may involve a $P$-wave version of polaron-to-molecule transition near the narrow Feshbach resonance~\cite{Massignan_2012}, which can occur even at negative scattering  volume~\cite{PhysRevLett.109.075302}.
Indeed, for the $S$-wave narrow Feshbach resonance,
it is known that the dressed
molecular state can be stabilized against the polaron formation even for negative $S$-wave scattering length.
In this case, the negative $S$-wave effective range plays a role.
While no numerical comparison between the polaron energy and the in-medium
$^5$He energy has been performed in Ref.~\cite{PhysRevC.106.045807},
it is obvious that such a transition can occur at sufficiently high densities where $1/(k_{\rm F}^3v_P)$ becomes close to $0$.

When the neutron density becomes of order 0.03 fm$^{-3}$,   
the neutron Fermi energy reaches the neutron separation energy of an alpha particle and hence
an alpha impurity is difficult to bind.
%may break up due to the medium effect.
Such a density is close to the so-called Mott density $\rho_{\rm Mott}$ at which  
an alpha-like cluster in bulk symmetric (neutron-proton equally populated) nuclear matter dissociates~\cite{PhysRevC.79.051301,PhysRevC.82.034322} just like a metal-insulator transition in strongly correlated electron systems.
In this regard, we can conclude that the present approach, which is based on point-like alpha particles, is qualitatively valid at neutron densities of up to $\simeq \rho_{\rm Mott}$.

\subsection{Polaronic protons}
While, in the preceding subsection, we have discussed the dressing of an alpha-like multi-nucleon cluster and its binding with one or two others as well as with a neutron
in a cold neutron gas, it is interesting to consider how a single proton behaves in such a gas.  In the dilute limit of the medium, the proton combines with an adjacent neutron to form a deuteron of binding energy of 2.2 MeV.  This deuteron, however, tends to dissociate once the medium density increases up to a density where $E_{\rm F}$ reaches the deuteron binding energy.
In such a case, a polaronic proton 
takes over~\cite{tajima2023polaronic}.
In neutron star matter,
the proton fraction
$Y_{\rm p}=\rho_{\rm p}/\rho$, where $\rho_{\rm p(n)}$ and $\rho=\rho_{\rm n}+\rho_{\rm p}$ are the number densities of protons (neutrons) and total nucleons, respectively, is 
basically around $Y_{\rm p}=0.01$--0.1, which is sufficiently small that protons can be regarded as impurities in neutron matter.
While the proton-proton interaction that has the Coulomb force subtracted out is almost the same as
the neutron-neutron one due to the charge symmetry of the strong interaction in vacuum,
this is not the case in the neutron medium, leading to the possible occurrence of diprotons as we shall see.

As in the case of polaronic alpha particles, 
a proton impurity, if moving slowly, is dressed by a virtual neutron particle-hole cloud and hence has its effective mass enhanced.
This cloud, however, is induced by the strong neutron-proton attraction, which is a contrast to the alpha impurity case in which the $S$-wave neutron-alpha repulsion acts to enhance the effective mass~\cite{PhysRevC.102.055802}.
In addition to such enhancement of the effective mass,
the single-particle energy, that is, the polaron energy $E_{\rm P}$, is lowered, 
as is the case with attractive Fermi polarons observed in cold atomic systems~\cite{PhysRevLett.102.230402,PhysRevLett.103.170402,PhysRevLett.118.083602}.
Since the nucleon-nucleon interaction involves a non-negligible magnitude of the
effective range~\cite{PhysRevC.51.38},
the polaron energy suffers
corrections due to the effective range
in such a way as to be consistent with the earlier studies~\cite{PhysRevC.103.L052801,PhysRevA.104.043313,PhysRevA.107.053313}.

Just as the ground-state energy density of the uniform system containing a nonzero density of atomic Fermi polarons can be expressed by Eq.~\eqref{eq:Epol}, 
%at small proton density $\rho_{\rm p}$ compared to the neutron one $\rho_{\rm n}$,
one can express, for $\rho_{\rm p}\ll\rho_{\rm n}$, the ground-state energy density $E$ of uniform nuclear matter in which the proton charge is assumed to be zero as
\begin{align}
\label{eq:Epnm}
    E=
    E_{\rm PNM}
    +E_{\rm P}\rho_{\rm p}
    +\frac{(3\pi^2)^{\frac{5}{3}}}{10\pi^2 M_{\rm p}^*}\rho_{\rm p}^{\frac{5}{3}}+\cdots
    %E_{\rm PNM}\left[1
    %+\frac{E_{\rm P}}{E_{\rm F,n}}\left(\frac{\rho_{\rm p}}{\rho_{\rm n}}\right)+\frac{M}{M^*}\left(\frac{\rho_{\rm p}}{\rho_{\rm n}}\right)^{\frac{5}{3}}+\cdots\right]
\end{align}
where 
$E_{\rm PNM}$ is the ground-state energy density of pure neutron matter, and $M_{\rm p}^*$ is the effective mass of a polaronic proton.
Indeed, Eq.~\eqref{eq:Epnm} is consistent with the energy expression of uniform nuclear
matter written
by Baym, Bethe, and Pethick~\cite{baym1971neutron}.
Most remarkably, they conjectured the enhancement of the effective mass $M_{\rm p}^*$ of an impurity proton
in Ref.~\cite{baym1971neutron}; this enhancement 
has just recently been confirmed by microscopic calculations within 
the strong-coupling polaron theory based on cold atom physics~\cite{tajima2023polaronic}.
In contrast to the single-impurity case, 
the so-called Landau effective mass is relevant for protons in uniform nuclear matter where the Fermi sphere
of protons is formed.  Even in neutron-rich nuclear matter at subnuclear densities
as encountered in neutron stars, the Landau effective mass of a proton quasiparticle on the Fermi surface
is predicted to be smaller than $M$~\cite{SJOBERG1976511,LI201829}.
To clarify how the polaronic and Landau effective masses are connected would be an interesting future work.

The form of $E$ in Eq.~(\ref{eq:Epnm}) can be compared with the well-known expansion
of the energy (per nucleon) of uniform nuclear matter from the isospin symmetric case ($\rho_{\rm n}=\rho_{\rm p}$) as given by~\cite{lattimer2012nuclear}
\begin{align}
\label{eq:Esnm}
    \frac{E}{\rho}=\frac{E_{\rm SNM}}{\rho}+S(\rho)
    \zeta^2+O(\zeta^4),
\end{align}
where $E_{\rm SNM}$ is the ground-state energy density of symmetric nuclear matter,
$S(\rho)$ is the symmetry energy,
and $\zeta=\left(\frac{\rho_{\rm n}-\rho_{\rm p}}{\rho}\right)$ is the isospin polarization. 
Here, the odd-order terms with respect to $\zeta$, which break the isospin symmetry, are ignored to a first good approximation.
Also, while the terms up to $\zeta^2$ are kept in Eq.~\eqref{eq:Esnm} for simplicity, microscopic calculations of the energy density at $0<\zeta<1$ are mostly favorable for such simplification, which motivates one to use Eq.~\eqref{eq:Esnm} at any $\zeta$.  Incidentally,
in Ref.\ \cite{baym1971neutron}, it was generalized
to the form with higher-order terms (i.e., $O(\zeta^4), O(\zeta^6),\cdots$).
By following the above simplification, one may rewrite Eq.~\eqref{eq:Esnm} as
\begin{align}
    \frac{E}{\rho}
    &=\frac{E_{\rm SNM}+\rho S(\rho)}{\rho}
    +S(\rho)(\zeta^2-1)+O(\zeta^4)\cr
        &=\frac{E_{\rm PNM}}{\rho}
    -4S(\rho)Y_{\rm p}+O(Y_{\rm p}^2),
\end{align}
where $E_{\rm PNM}=E_{\rm SNM}+\rho S(\rho)$ corresponding to the limit of $\zeta\rightarrow 1$ in Eq.~\eqref{eq:Esnm}. 
In this regard, focusing on the term proportional to $\rho_{\rm p}/\rho$,
one can relate $E_{\rm P}$ to $S(\rho)$ as
\begin{align}
    S(\rho)=-\frac{1}{4}E_{\rm P}%\left(\frac{\rho}{\rho_{\rm n}}\right)
    +O(Y_{\rm p}),
    %\simeq -\frac{1}{2}E_{\rm P} \quad (\rho\rightarrow \rho_{\rm n}),
\end{align}
which indicates that the determination of $E_{\rm P}$ can be a new route to address the symmetry energy from 
a neutron gas.
Furthermore, $S(\rho)$ can be expanded around $\rho=\rho_0$
%the normal nuclear density $\rho_0\simeq 0.16$ fm$^{-3}$ 
as~\cite{PhysRevC.75.015801}
\begin{align}
    S(\rho)\simeq S_0+\frac{L_0}{3\rho_0}(\rho-\rho_0),
\end{align}
where $S_0\equiv S(\rho=\rho_0)$ and $L_0$ is called the slope parameter.
This implies that $L_0$ can also be extracted from the dependence of $E_{\rm P}$ on the medium density.
$L_0$, which characterizes the stiffness of the equation of state of neutron-rich matter, remains to be well determined, although
various nuclear experiments have been performed to determine $L_0$~\cite{doi:10.1080/10619127.2017.1388681}.
The approach based on $E_{\rm P}$ can address $L_0$
from pure neutron matter, i.e.,
in the opposite direction to the usual approach. 
Future experimental study on protons in the neutron skin region of neutron-rich nuclei might give us information about polaronic protons.

Another interesting aspect of polaronic protons is the neutron-mediated proton-proton interaction leading to the formation of bound diprotons, i.e., $^2$He nuclei, which are known to be unbound in vacuum (see also Fig.~\ref{fig:2}).
As long as we can regard protons as impurities in a neutron gas,
the neutron-proton pairing is strongly suppressed by the Fermi-surface mismatch between neutrons and 
protons~\cite{PhysRevC.64.064314,tajima2019superfluid}.
The strong neutron-proton coupling is nevertheless effective at inducing a strong proton-proton attraction in addition to the
in-vacuum attraction in the $S$-wave spin-singlet channel.
Such an additional attraction has
been studied in connection with the medium polarization~\cite{PhysRevC.93.044329}, which is analogous to the RKKY-type polaron-polaron interaction in the case of polaronic alpha particles (see the previous subsection) and cold atomic polarons.
While, in a low-density neutron gas, spin fluctuations, which dominate
the medium polarization, act to weaken the neutron-neutron attraction in the $S$-wave spin-singlet channel
and hence decrease the neutron pairing gap~\cite{PhysRevC.74.064301,PhysRevC.67.061302,PhysRevC.101.035803},
the neutron-mediated proton-proton interaction is attractive
because neutron density fluctuations, which work between two protons with anti-parallel spins, dominate the medium polarization.
Such an interaction, together with the enhanced polaron effective mass, may induce proton clustering in the neutron medium in the form of 
diprotons or even heavier nuclei that
are unbound in vacuum.

%\textbf{Ethical approval declarations} (only required where applicable) Any article reporting experiment/s carried out on (i)~live vertebrate (or higher invertebrates), (ii)~humans or (iii)~human samples must include an unambiguous statement within the methods section that meets the following requirements: 

%\begin{enumerate}[1.]
%\item Approval: a statement which confirms that all experimental protocols were approved by a named institutional and/or licensing committee. Please identify the approving body in the methods section

%\item Accordance: a statement explicitly saying that the methods were carried out in accordance with the relevant guidelines and regulations

%\item Informed consent (for experiments involving humans or human tissue samples): include a statement confirming that informed consent was obtained from all participants and/or their legal guardian/s
%\end{enumerate}

%If your manuscript includes potentially identifying patient/participant information, or if it describes human transplantation research, or if it reports results of a clinical trial then  additional information will be required. Please visit (\url{https://www.nature.com/nature-research/editorial-policies}) for Nature Portfolio journals, (\url{https://www.springer.com/gp/authors-editors/journal-author/journal-author-helpdesk/publishing-ethics/14214}) for Springer Nature journals, or (\url{https://www.biomedcentral.com/getpublished/editorial-policies\#ethics+and+consent}) for BMC.

\section{Conclusion}\label{sec:4}
In this article, we have given an overview of cold atom quantum simulation of nuclear impurity problems, which has been theoretically developed by us for the past few years to understand the poorly known properties of extremely neutron-rich nuclear matter as encountered in neutron stars as well as in the neutron skin region of neutron-rich nuclei.
The most important prediction is that some of the light nuclides unbound in vacuum can be bound in a cold neutron gas,
as summarized in Fig.~\ref{fig:2}.
This prediction is based on the polaron picture, which is well established for extremely population-imbalanced cold atoms. 
In the course of the present study, we have also found that the polaron energy of an impurity proton in neutron matter could give a novel piece of information about the dependence of the symmetry energy on the nucleon density, which is one of the main topics of research in nuclear physics.

To make better estimates of the nuclear polaronic properties, however, many problems remain.
While we have discussed nuclear impurity problems by assuming that the neutron medium is in a normal gas state, at sufficiently low temperatures, the neutron gas generally undergoes the superfluid phase transition.
In this case, one needs to consider the effects of the neutron superfluidity as done in the context of ultracold Fermi gases~\cite{PhysRevLett.114.115302,PhysRevA.92.013620,PhysRevA.105.023317}.
For applications to neutron star mergers and supernova explosions, finite temperature effects on Fermi polarons as experimentally clarified for cold atomic impurities~\cite{PhysRevLett.122.093401}, have to be allowed for, especially for temperatures of order or even higher than the neutron Fermi temperature.
It is also interesting to see what a system of many polaronic protons ends up with in a neutron gas, a question relevant to the structure of neutron star matter at subnuclear densities where the liquid part could be modified by the medium-induced interaction between polaronic protons.
Another interesting open question is how empirically known bound nuclei such as $^9$Be, which are considered to be composed of alpha-like clusters and excess neutrons, can be understood in terms of polaronic alpha particles.
These are left for future work.

%Conclusions may be used to restate your hypothesis or research question, restate your major findings, explain the relevance and the added value of your work, highlight any limitations of your study, describe future directions for research and recommendations. 

%In some disciplines use of Discussion or 'Conclusion' is interchangeable. It is not mandatory to use both. Please refer to Journal-level guidance for any specific requirements. 

\backmatter

\bmhead{Acknowledgments}
This work was supported by Grants-in-Aid for Scientific
Research provided by JSPS through Nos.~18H05406, 18K03635, 21K03422, 22H01158, 22H01214, 22K13981, and 23H01167.

\section*{Declarations}
Not applicable.

\bibliography{sn-bibliography}% common bib file


\end{document}